\documentstyle[12pt]{article}

\date{}

\begin{document}
\title{Effects of Nonlinearity in Wave Propagation \\ 
in Multicomponent Bose--Einstein Condensates}
\author{I.E.Mazets$^1$\, \footnote{E-mail: mazets@astro.ioffe.rssi.ru}
, E.V.Orlenko$^2$, B.G.Matisov$^2$,\\
$^1${\normalsize 
{\it A.F.Ioffe Physico-Technical Institute, 194021 St.Petersburg, 
Russia}}\\
$^2${\normalsize{\it St.Petersburg State Technical University, 195251 
St.Petersburg, Russia}}}
\maketitle

\begin{abstract}
We consider a spinor Bose-Einstein condensate 
in its polar ground state. We analyze magnetization waves of a finite 
amplitude and show that their nonlinear coupling to the density waves 
change the dependence of the frequency on the wavenumber dramatically. 
In contrary, the density wave propagation is much less modified by the 
nonlinearity effects. A similar phenomenon in a miscible 
two-component condensate is studied, too. 
\end{abstract}



Recent advances in experimental creation of multicomponent atomic 
Bose--Einstein condensates \cite{jn,mn,ob} have given rise to an 
interest to physical properties of such systems. There are numerous 
works on the properties of degenerate Bose gas mixtures in 
magnetic traps related to both the ground state \cite{gsmx} and the 
collective excitations \cite{tt}. In Ref.\cite{tt} the early work 
\cite{nep} related to a homogeneous Bose gas mixture is generalized to 
the case of presence of external harmonic trap potential. Of course, 
the number of branches of the dispersion law is equal to the number of 
different components in a mixed BEC. Due to non-zero interaction 
between them, normal mode oscillations imply 
simultaneous mutually coherent motion of the components. 
In the present paper, we, however, consider firstly a multicomponent BEC 
of another kind, namely, a spinor BEC. Such a degenerate quantum system 
can be created in an optical trap, where all the atoms are confined 
practically independently on $m_f$, their momentum projection to any 
arbitrary axis. Such an independence of confinement on the spin 
orientation is a striking feature and a key advantage of an optical trap, 
well justified experimentally \cite{mn,ob}. Under such a condition, 
the spin orientation becomes a new degree of freedom. The differences 
and similarities between a two-component BEC with fixed $m_f$'s for both 
the components and a spinor BEC in context of our study 
will be discussed later, at the end of this paper. 

Now we have to note that in all the cited works on collective 
excitations in multicomponent BECs as well as in the seminal 
works on spinor BEC dynamics \cite{fod} oscillation amplitudes were 
assumed to be small enough to provide linearization of the set of 
coupled time-dependent Gross--Pitaevskii equations (GPEs). Then a proper 
linear transformation yields the equations of the harmonic oscillator 
type for the normal modes. However, the GPE is essentially nonlinear, 
thus the effects of finite amplitude of oscillations take place. 
There are some approaches to take nonlinearity into account. The first one 
is to find particular solutions of the GPE in a form of solitons (see, 
e.g., the recent work \cite{so} and references therein). The second one is 
to find oscillating nonlinear solutions those in the case of 
infinitesimally small oscillation amplitude coincide with corresponding 
eigenfunctions of the linearized version of the GPE or of the equivalent  
set of quantum hydrodynamical equations. An elegant formalism has been 
developed for nonlinear oscillations of a scalar BEC in a harmonic trap 
in the Thomas--Fermi regime \cite{csn}. It has been found that the 
nonlinear effects become important, if the fraction of mass of a scalar 
BEC involved into oscillatory motion is comparable to unity. 

In the present paper, we study validity of the approximation based on 
linearization of the GPE 
by proceeding in the following way. We consider 
plane waves in a spatially homogeneous multicomponent BEC. This 
can serve as a WKB approximation for excitations in a trapped BEC, if 
the excitation wavelength is much smaller than the atomic cloud size. 
Moreover, such an approach allows us to use in the most direct and 
straightforward way the standard technique of expanding a solution in 
series in a certain small parameter, known as a standard 
perturbation theory in classical mechanics \cite{ptm}. Also consideration 
of plane waves in a translationally invariant BEC provides a possibility 
of comparing the results to the strict analytic formulae of 
Refs.\cite{nep,fod}. 

The main result of our work is that certain modes in a multicomponent 
BEC exhibit strongly nonlinear behaviour, namely, even for a relatively 
small wave amplitude the effects of anharmonicity become significant. 
This effect is absent in a scalar BEC case. 

Let us consider a spinor BEC composed of atoms with the spin $f=1$ 
at zero temperature. The 
GPE governing evolution of the complex order parameter (macroscopic 
wave function) 
$\psi ({\bf r},\,t)$ of the BEC reads in the mean field approximation 
as follows \cite{fod}: 
\begin{equation}
i\hbar \frac \partial {\partial t}\psi =-\frac {\hbar ^2}{2M}\nabla ^2\psi 
-\mu \psi +\hbar c_0(\tilde \psi ^*\psi )\psi +\hbar c_2(\tilde \psi ^* 
\hat{\bf f}\psi )\cdot (\hat{\bf f}\psi ), 
\label{spmf}
\end{equation} 
where $\hat{\bf f}$ is the single-atom angular momentum operator, a 
vector with Cartesian components being $3\times 3$ matrices, $M$ is the 
mass of an atom, $\mu $ is the chemical potential. Interaction constants 
defined as $\hbar c_0=(g_0+2g_2)/3$, $\hbar c_2=(g_2-g_0)/3$, 
$g_F=4\pi \hbar^2a_F/M$, and $a_F$ is the $s$-wave scattering length for 
a pair of slow atoms with the total angular momentum $F$ equal to 0 or 2, 
respectively. Practically, the magnitudes of these two scattering 
lengths are close each to other, so $\left| c_2/c_0\right| \ll 1$. 
The order parameter $\psi $ has three components, corresponding to the 
momentum projection to the $z$-axis $m_f=0,\,\pm 1$: 
$$
\psi =\left( 
\begin{array}{l} \zeta _1\\ \zeta _0\\ \zeta _{-1}\end{array} \right) 
\sqrt{n}, 
$$
where $n$ is the total equilibrium density of the BEC. $\tilde \psi $ 
means the transposed vector. In other words, the ground state components 
of the vector $\zeta $ are normalized by the condition 
\begin{equation}
\sum _{m_f=-1}^1 \left| \zeta _{m_f}^{(ground)}\right| ^2=1. 
\label{hd}
\end{equation} 

We assume the repulsive interaction of atoms in the BEC, 
i.e., $c_0>0$. For the sake of definiteness, we assume also that 
$c_2>0$. It follows from the latter condition that the ground state of 
such a system is a so-called polar state \cite{fod}. 
This means that, in the 
mean field picture, all the atoms have zero momentum projection to a 
certain axis. This state is degenerate with respect to orientation 
of this axis. Let this axis be the $z$-axis, so in the equilibrium, 
when the time derivative of $\psi $ in Eq.(\ref{spmf}) is equal to zero, 
$$\zeta _{\pm 1}^{(ground)}=0, \qquad \zeta _0^{(ground)}=1.$$ 
The chemical potential of the BEC in the polar state is $\mu =c_0n$. 

Before writing Eq.(\ref{spmf}) in explicit form, we introduce the 
new unknown functions: $\xi _\pm =(\zeta _1\pm \zeta _{-1}^*)/\sqrt{2}$, 
$\eta _p=$~Re$\,\zeta _0-1$, $\eta _i=$~Im$\,\zeta _0$. Then 
Eq.(\ref{spmf}) can be transformed to the set of equations 
\begin{eqnarray}
-\frac \partial {\partial t}\xi _-&=&-\frac \hbar {2M}\nabla ^2\xi _+ +
2c_2n\xi _+ +c_0n(\xi _+^*\xi _++\xi _-^*\xi _-+2\eta _p+\eta _p^2 
+\eta _i^2)\xi _+ +
\nonumber \\ & & c_2n[(\xi _+\xi _-^* - \xi _+^*\xi _- + 
2\eta _i+2\eta _p\eta _i)\xi _- +2(2\eta _p+\eta _p^2)\xi _+] ,
\label{xe1} \end{eqnarray}
\begin{eqnarray} 
\frac \partial {\partial t}\xi _+&=&-\frac \hbar {2M}\nabla ^2\xi _- +
c_0n(\xi _+^*\xi _++\xi _-^*\xi _-+2\eta _p+\eta _p^2 
+\eta _i^2)\xi _- +\qquad \quad \nonumber \\ & &
c_2n[(\xi _+\xi _-^* - \xi _+^*\xi _- + 
2\eta _i+2\eta _p\eta _i)\xi _+ +2\eta _i^2\xi _-],
\label{xe2} \end{eqnarray}
\begin{eqnarray}
-\frac \partial {\partial t}\eta _i&=&-\frac \hbar {2M}\nabla ^2\eta _p+
2c_0n\eta _p+c_0n[(\xi _+^*\xi _+ + \xi _-^*\xi _-+3\eta _p+
\eta _p^2+\eta _i^2)\eta _p+\nonumber   \\ & & \xi_+^*\xi _+ +
\xi_-^*\xi _-+\eta _i^2] +c_2n[2\xi _+^*\xi _+\eta _p +2\xi _+^*\xi _+ +
(\xi_+^*\xi_-+\xi_+\xi _-^*)\eta _i], 
\label{xe3}   \end{eqnarray} 
\begin{eqnarray}
\frac \partial {\partial t}\eta _p&=&-\frac \hbar {2M}\nabla ^2\eta _i+
c_0n(\xi _+^*\xi _+ +\xi _-^*\xi _-+2\eta _p+\eta _p^2+\eta _i^2)\eta _i+
\qquad \quad \nonumber \\ & & c_2n[2\xi _-^*\xi _-\eta _i+(\xi_+^*\xi_- + 
\xi _+\xi _-^*)\eta _p+\xi _+^*\xi _- +\xi _+\xi _-^*]. 
\label{xe4}          \end{eqnarray}
If we neglect all the nonlinear terms in Eqs.(\ref{xe1}\,---\,\ref{xe4}), 
then we get immediately the solutions in the form of plane 
monochromatic waves and the corresponding dispersion laws 
\cite{fod}. The first mode is the density wave, it corresponds, 
in the linear approximation, to the periodic 
oscillations of the $m_f=0$ component 
of the order parameter only (i.e., of $\eta _p,\, \eta _i$), while 
$\xi _+$ and $\xi _-$ remain zero. Density waves in a spinor BEC 
are the same as sound waves in a scalar BEC. 
The dependence of the frequency 
$\omega _{d0}$ of the density waves on the wavenumber $k$ is of the 
Bogoliubov's type, $\omega _{d0}^2(k)=\omega _r(k)[\omega _r(k)+2c_0n]$, 
where $\omega _r(k)=\hbar k^2/(2M)$ is the recoil frequency associated 
with the kinetic momentum $\hbar k$. Another branch of the excitation 
spectrum in a spinor BEC is related to magnetization waves. Left and 
right circularly polarized magnetization modes are degenerate and 
in the linear regime their frequency is given by the formula 
$\omega _{m0}^2(k)=\omega _r(k)[\omega _r(k)+2c_2n]$. The quantum 
mechanical mean values of the atomic magnetic momentum operator are 
proportional to the $\xi _+$ and $\xi _+^*$ for the left and right 
polarization, respectively. 

Now we can determine the effects of nonlinearity on magnetization 
wave propagation, using the perturbation theory of classical mechanics 
\cite{ptm}. Namely, we expand our unknown functions in series: 
$\xi _+=\sum _{j=0}^\infty \xi _+^{(j)}$, where $\xi _+^{(j)}$ is 
proportional to the $j$-th power of a certain small parameter 
$\varepsilon $ (in fact, the square of the magnetization amplitude can 
be naturally regarded as this parameter). 
Similar expansions hold for the remaining three functions. 
The zeroth order approximation can be taken also in the form of the 
plane wave, $\xi _+^{(0)}=A_+\sin \,(\omega t-{\bf k\,r})$, but with the 
frequency $\omega $ shifted with respect to the non-perturbed value 
$\omega _{m0}$. The validity of this method is restricted to the case 
of small resulting correction to the frequency, $\left| (\omega -
\omega _{m0})/\omega _{m0} \right| \ll 1$. 
Also we take $\xi _-^{(0)}=\omega _r(k)^{-1}\omega A_+
\cos \, (\omega t-{\bf k\,r})$, $\eta_p^{(0)}=0$, $\eta_i^{(0)}=0$. 
The difference between $\omega $ and $\omega _{m0}$ can be also 
represented as a series in $\varepsilon $, beginning from the term 
of order of $\varepsilon ^1$. 

To find the correction to the frequency of a 
magnetization wave, we make the 
following transformation of our set of GPEs. 
We add to and substract from the right-hand side of Eq.(\ref{xe1}) 
the term $\omega ^2\xi _+/\omega _r(k)$. Then we note that our zeroth  
order approximation satisfies the set of equations $-\partial \xi_-/
\partial t=\omega ^2\xi _+/\omega _r(k)$, $\partial \xi _+/\partial t=
\omega _r(k)\xi _-$ identically. The remaining terms must be regarded as 
a perturbation leading to the frequency shift in higher orders of 
approximation. Eqs.(\ref{xe1}\,---\,\ref{xe4}) must be satisfied in every 
order in $\varepsilon $ separately, i.e., one must group all the terms 
of order of $\varepsilon ^j$ in the right-hand side and equalize them to 
the $O(\varepsilon ^j) $ part of the left-hand side of the equation.  
We restrict our analysis to the linear order in $\varepsilon $ when we 
obtain 
\begin{eqnarray}
-\frac \partial {\partial t}\xi _-^{(1)}&=&
\frac {\omega ^2}{\omega _r(k)}\xi_+^{(1)}+\Big{\{ }
\omega _r(k)+2c_2n-\frac {\omega ^2}{\omega _r(k)}\Big{\} }^{(1)} 
A_+\sin \,(\omega t-{\bf k\,r})+\qquad \nonumber \\ & & 
c_0n\Big{[} \sin ^2\,(\omega t -{\bf k\,r})+
\frac {\omega ^2}{\omega _r^2(k)}
\cos ^2\,(\omega t-{\bf{k\, r}} )\Big{]}
A_+^3\sin \,(\omega t-{\bf k\,r}),
\label{x1a0} \end{eqnarray}
\begin{eqnarray} 
\frac \partial {\partial t}\xi_+^{(1)}&=&\omega_r(k)\xi_-^{(1)}+
\frac {\omega }{\omega _r(k)}
c_0n\Big{[} \sin ^2\,(\omega t -{\bf k\,r})+
\frac {\omega ^2}{\omega _r^2(k)}\cos ^2\,(\omega t-{\bf k\, r} )\Big{]}  
\times \nonumber \\ & & 
A_+^3\cos\, (\omega t-{\bf{k\,r}}). 
\label{x1b0}           \end{eqnarray}
Here the symbol $\{ \, ...\, \}^{(1)}$ means that only linear 
in $\varepsilon \sim A_+^2$ contribution to the expression in the curly 
brackets is retained. $A_+$ is taken to be real, without loss 
of generality. 

Eqs.(\ref{x1a0}, \ref{x1b0}) can be easily reduced to the following 
differential equation 
\begin{eqnarray}
\frac {\partial ^2}{\partial t^2}\xi _+^{(1)} +\omega ^2\xi_+^{(1)}+
\Big{\{ }\omega _{m0}^2-\omega ^2+\frac {c_0n\omega _r(k)}4\Big{[}3+
4\frac {\omega ^2}{\omega _r^2(k)}+ 
3\frac {\omega ^4}{\omega _r^4(k)}\Big{]}A_+^2\Big{\} }^{(1)} 
& \times & \nonumber \\ 
A_+\sin \,(\omega t-{\bf k\,r})+
CA_+^3\sin \,[3(\omega t-{\bf k\,r})] & = & 0. 
\label{x11}                         \end{eqnarray} 
Here $C$ is a certain combination of various frequency parameters of 
the problem; its calculation is not needed for determination of 
the correction to the wave frequency in the lowest order.

Eq.(\ref{x11}) is inhomogeneous, and the presence of resonant source 
term proportional to $\sin \,(\omega t-{\bf k\,r})$ leads to 
occurrence of oscillations with amplitude growing linearly in time 
in the solution for $\xi _+^{(1)} $. And the very essence of the 
method used here \cite{ptm} is avoiding of these non-physical 
(secular) solutions proportional to $t\sin (\omega t-{\bf k\,r})$ 
by setting the prefactor of the resonant term 
to zero. Thus, to the lowest order in square of the wave amplitude, the 
magnetization wave frequency is given by the expression 
\begin{equation}
\omega ^2 =\omega _{m0}^2+\frac {c_0n\omega _r(k)}4 \Big{[} 3+
4\frac {\omega _{m0}^2}{\omega _r^2(k)}+
3\frac {\omega _{m0}^4}{\omega _r^4(k)} \Big{]}A_+ ^2. 
\label{mwp}
\end{equation}
In the two limiting cases (of the short and long wavelength) we obtain 
\begin{equation}
\omega ^2=\omega _r^2(k)+\frac 54 u_0^2k^2A_+^2, \qquad \hbar k\gg Mu_2, 
\label{mwps} 
\end{equation}
and
\begin{equation}
\omega ^2=u_2^2k^2+6u_0^2k^2\left( \frac {Mu_2}{\hbar k}\right) ^4A_+^2,  
\qquad \hbar k\ll Mu_2, 
\label{mwpl}
\end{equation}
respectively. Here $u_F=\sqrt{\hbar c_F n/M}$ are the velocities of 
propagation of density ($F=0$) and magnetization ($F=2$) waves of 
infinitely small amplitude in the long wavelength limit. 
So we can conclude that the effects of nonlinearity are small until 
\begin{eqnarray}
A_+^2\ll  1, &\qquad &\hbar k\gg Mu_0,             \label{scm1}
\\ 
A_+^2\ll \left( \frac {\hbar k}{Mu_0}\right) ^2, &\qquad &
Mu_2{~ ^<_\sim ~} \hbar k{~ ^<_\sim ~} Mu_0,   \label{scm2}
\\
A_+^2 \ll \frac {c_2}{c_0}\left( \frac {\hbar k}{Mu_2}\right) ^4 , 
&\qquad &\hbar k\ll Mu_2.                      \label{scm3}
\end{eqnarray}    
This is interesting that the condition of the nonlinearity smallness 
coincides with the trivial condition of smallness of $A_+^2$ in 
comparison to the sum of squares of absolute values of all the three 
$\zeta ^{(ground)}_{m_f}$ in the ground state which is unity, 
according to Eq.(\ref{hd}), only in the short wavelength limit 
of Eq.(\ref{scm1}). 
In the other other cases [Eqs.(\ref{scm2}, \ref{scm3})], even small 
but finite excitation amplitude can result in  a significant 
modification of the wave propagation. 

It is easy to show that 
Eqs.(\ref{xe3}, \ref{xe4}), in the case of magnetization waves, have 
no resonant terms in their right-hand sides in the first order in 
$\varepsilon $ and hence do not contribute to the evaluation of 
the corresponding correction to the wave frequency. 

Density waves can be analyzed in the similar way, and the lowest-order 
correction results in the formula 
\begin{equation}
\omega ^2=\omega _{d0}^2+\frac 34c_0n\omega _r(k)A_p^2, 
\label{sc6}
\end{equation}
where $A_p$ is the amplitude of oscillations of $\eta _p$. For all the 
momenta $k$, the correction is small provided that $A_p\ll 1$, i.e., 
nonlinear effects play less role for waves of this type, in contrary to 
magnetization waves. Eq.(\ref{sc6}) also applies to sound waves in a 
single-component (scalar) BEC.        

Since $\omega _{m0}$ does not depend on $c_0$ but the latter quantity 
appears in the right-hand side of Eq.(\ref{mwp}), we conclude that 
nonlinear coupling to density waves plays the key role in 
modification of the magnetization wave frequency. In opposite, 
Eq.(\ref{sc6}) does not contain $c_2$, so a traveling density 
wave is not coupled to magnetization modes. 

Now let us discuss briefly the case of a mixture of two BECs each 
having the fixed value of $m_f$ or, equivalently, of two scalar BECs. 
Here we need first to introduce the coupling constants 
$g_{j^\prime j}=2\pi \hbar a_{j^\prime j}(M_j+M_{j^\prime })/
(M_jM_{j^\prime })$, where $M_j$ is the mass of an atom of the $j$-th 
kind, $a_{j^\prime j}$ is the $s$-wave scattering length for a pair 
of atoms of $j$-th and $j^\prime $-th kind, $j^\prime,\, j=1,\, 2$. 
The dispersion laws for the two excitation branches were obtained 
in the analytic form in Ref.\cite{nep} (see also Ref.\cite{tt}). 
If all the three relevant scattering lengths are positive, the 
criterion of stability of a homogeneous BEC mixture against phase 
separation is simply $g_{12}<\sqrt{g_{11}g_{22}}$. In this case, the 
eigenmode frequencies are positive for all the values of the momentum $k$. 
For the sake of simplicity, we consider in our paper the case of 
equal atomic masses, $M_1=M_2\equiv M$. Then the eigenfrequencies 
are simply $\omega _\pm^2=\omega _r(k)[\omega _r(k)+2\Lambda _\pm ]$,  
where $\Lambda _\pm =[g_{11}n_1+g_{22}n_2\pm 
\sqrt{(g_{11}n_1-g_{22}n_2)^2+4g_{12}^2n_1n_2}\, ]/2$, $n_1,\, n_2$ are 
the equilibrium number densities of the components, $\omega _r(k)$ is 
the same quantity as defined above. 

The order parameter perturbation for the $j$-th component reads as 
$\delta \psi _j=\sqrt{n_1}A_j[\sin (\omega t-{\bf k\,r})+
i\omega _r^{-1}(k)\omega \cos(\omega t-{\bf k\,r})]$. After some 
tedious but straightforward calculations, analogous to those described 
above and valid under the same condition of smallness of the 
frequency correction, 
we arrive at the following formula for the shifted, due to 
the non-linearity effects, wave frequency: 
\begin{equation}
\omega ^2=\omega _\pm ^2+ 
\frac {\omega _r(k)g_\pm n_1}2\Big{[}3+4\frac {\omega _\pm ^2}
{\omega _r^2(k)}+3\frac {\omega _\pm ^4}{\omega _r^4(k)}\Big{]}
B_\pm ^2. 
\label{n11}
\end{equation}
Here the upper sign corresponds to the case of $B_+\ne 0$, $B_-=0$, and 
the lower sign corresponds to the opposite case, $B_+=0$, $B_-\ne 0$. 
Here the eigenmode amplitudes are defined as 
\begin{equation}
B_+= \cos \theta _g\,A_1+\sqrt{\frac {n_2}{n_1}}\sin \theta _g\,A_2
,\qquad 
B_-=-\sin \theta _g\,A_1+\sqrt{\frac {n_2}{n_1}}\cos \theta _g\,A_2. 
\label{AA5}
\end{equation}
Also we set by definition 
\begin{eqnarray}
g_+&=&g_{11}\cos ^4\theta _g+2g_{12}\cos ^2\theta _g\,\sin ^2\theta _g+
g_{22}\sin ^4\theta _g,                              \label{Gp} \\
g_-&=&g_{11}\sin ^4\theta _g+g_{22}\cos ^4\theta _g, \label{Gm}  
\end{eqnarray}
\begin{equation}
\tan \theta _g=\frac {g_{22}n_2-g_{11}n_1+\sqrt{(g_{22}n_2-g_{11}n_1)^2+
4g_{12}^2n_1n_2}}{2g_{12}\sqrt{n_1n_2}}.             \label{Qg}
\end{equation} 

Eq.(\ref{n11}) is of the form similar to Eq.(\ref{mwp}) and leads to 
similar restriction on the wave amplitude. If we the two BECs are 
composed of atoms accumulated on two different 
magnetic or hyperfine sublevels of the ground internal state, the 
difference between $g_{12}$ and $\sqrt{g_{11}g_{22}}$ is relatively 
small, and the lower-frequency mode is extremely sensitive to the 
effects of nonlinearity in the long wavelength limit. One should note 
that both the branches of excitation spectrum of a two-component BEC in 
an external magnetic field are sensitive to nonlinear effects for small 
$k$'s, while the spinor BEC collective excitations exhibit different 
behaviour: the nonlinearity effects are important for magnetization 
waves much more than for density waves. 

In summary, we should note that the studied effects of nonlinearity 
in wave propagation in a BEC are related to the Beliaev damping \cite{zb}
(cf. the closely related recent publication \cite{ac} on an 
efficient damping of a relative motion of two condensates in a trap 
by nonlinear interaction). 
The Beliaev damping is also described the cubic nonlinear term in 
the GPE. It is, in fact, decay of a quantum of collective excitation to 
two quata of lower energies, provided that the energy and momentum are 
conserved. This process results in occurrence of an imaginary part of 
the wave frequency (the damping constant). We calculate in the present  
paper the real small addend to the wave frequency. While the Beliaev 
damping becomes less important when $k$ approaches zero, the 
nonlinear corrections to the magnetization mode in a spinor BEC and to 
both of the modes in an usual two-component BEC become more pronounced. 

Finally, we present a numerical example. 
The ground state of a spinor BEC of sodium atoms with $f=1$ is just 
a polar (antiferromagnetic) state \cite{mn}. We take  
$(a_0+2a_2)/3\approx 5$~nm, $(a_2-a_0)/3\approx 0.08$~nm and set  
$n\approx 10^{14}$~cm$^{-3}$. Let the excitation wavenumber be of 
about $3.5\cdot 10^3$~cm$^{-1}$ (the corresponding wavelength is 
several times smaller than the atomic cloud size in an experiment with 
large number of atoms in a trap like that of Ref.\cite{mn}, so the WKB 
approximation still remains satisfactory). For $A_+\rightarrow 0$ the 
linear theory \cite{fod} gives the magnetization wave frequency 
$\omega _{m0}\approx 300$~s$^{-1}$. However, if $A_+\approx 0.044$, in 
other words, only $[1+\omega _r^{-2}(k)\omega ^2]A_+^2/2\approx 0.005$ 
of the total mass of the BEC is involved into the motion, then the 
frequency rises by one-third of its primary value and becomes equal to 
400~s$^{-1}$, according to Eq.(\ref{mwp}). Similarly, one can expect 
a strongly nonlinear behaviour of low-lying magnetization modes of a 
spinor BEC in an optical trap of a finite size, since the trapped BEC 
spectrum ought to reveal the most important qualitative features 
present in the translationally invariant case, as it has been shown 
for two-component BECs in magnetic traps \cite{tt}. 

This work is supported by the NWO, project NWO--047--009.010, the state 
program ''Universities of Russia'', grant 015.01.01.04, and the 
Ministry of Education of Russia, grant E00--3--12.

\end{document}